# Thermal conductivity of Magnesium Telluride (MgTe) - A first principles study

Rajmohan Muthaiah, Jivtesh Garg

School of Aerospace and Mechanical Engineering, University of Oklahoma, Norman,

OK-73019, USA

**Abstract:** In this work, we report thermal conductivity($k$) of magnesium telluride (MgTe) with various crystallographic phases such as rocksalt, zincblende, wurtzite and nickel arsenic (NiAs) using density functional theory and Boltzmann transport equation. Our first principles calculations results show the low thermal conductivity of MgTe with $k_{NiAs} < k_{rocksalt} < k_{wurtzite} < k_{zincblende}$. We systematically investigated the phonon group velocity, phonon scattering rate and mode contributed thermal conductivity of transverse acoustic (TA), longitudinal acoustic (LA) and optical phonons. Our first principles calculations shows that ultra-low thermal conductivity of 2.645 Wm$^{-1}$K$^{-1}$ for NiAs phase is due to the dominant scattering of TA and LA phonons by low frequency optical phonons. We also analyzed the length dependence thermal conductivity of MgTe at nanometer length-scales. At nanometer length scales such as 50 nm for NiAs phase, room temperature thermal conductivity of less than 1.4 Wm$^{-1}$K$^{-1}$ shows a promising nature of MgTe for thermoelectric applications.

**Introduction:** Magnesium chalcogenides-based semiconductors have attracted both scientific and technological applications[1-3]. Magnesium[4-7] and Telluride[8-14] based thermoelectric and photovoltaic materials are getting attention among the scientific community due to its ultra-low thermal conductivity and tunable electronic bandgap. Magnesium telluride(MgTe) is extensively studied for its structural[1, 2, 15, 16], electronic[1-3, 15], elastic[15], magnetic[1, 2], optical[15] and vibrational[3, 17] properties. Despite these extensive studies, thermal conductivities of MgTe are unknown and inspiring us to compute it for all the crystalline phases. Thermal conductivity of a material is critical for wide varieties of application such as thermal management system[18-25], thermoelectrics[26-28], opto-electronics[29] and solar cells[30, 31] etc., MgTe are known to exist in four crystalline phases such as zinc-blende (ZB), rocksalt (RS), wurtzite (WZ)[3, 32, 33] and nickel arsenic (NiAs). In this work, we report bulk and nanoscale thermal conductivity of all the four crystalline phases of MgTe using density functional theory and phonon Boltzmann transport equation. We also report an ultra-low thermal conductivity of MgTe at nanometer length scales. We systematically investigated the elastic constants, phonon group velocity, phonon bandgap and phonon scattering rate (inverse of phonon lifetime) for all the crystalline phases. At 300 K, bulk thermal conductivity of 2.645(NiAs), 6.26(RS), 8.83(WZ) and 10.05(ZB) Wm$^{-1}$K$^{-1}$ shows that MgTe will be a promising thermoelectric material. These results have important implications for applications of MgTe in thermoelectric energy conversion techniques, solar-cells and other opto-electronics.

**Computational Methods:**

All the first principles calculations were performed using QUANTUM ESPRESSO[34] package. Norm-conserving pseudopotential with local density approximation (LDA)[35] exchange-correlation functional is used to approximate the MgTe. The geometry of the zinc-blende and rocksalt MgTe with 2 atoms (4 atoms for wurtzite and NiAs) unit cell were optimized until forces on all atoms were less than 10$^{-6}$ Ry/bohr. Plane-wave energy cutoff of 80 Ry and 8 x 8 x 8(12 x 12 x 8) Monkhorst-Pack[36] $k$-point mesh were used integrate over the Brillouin zone. Relaxed structure with equilibrium lattice constants of MgTe with different lattice crystal phases are shown in Fig 1 and also listed in Table 1 which are in excellent agreement with previously published values G.P.Srivatsava *et.al*[3].

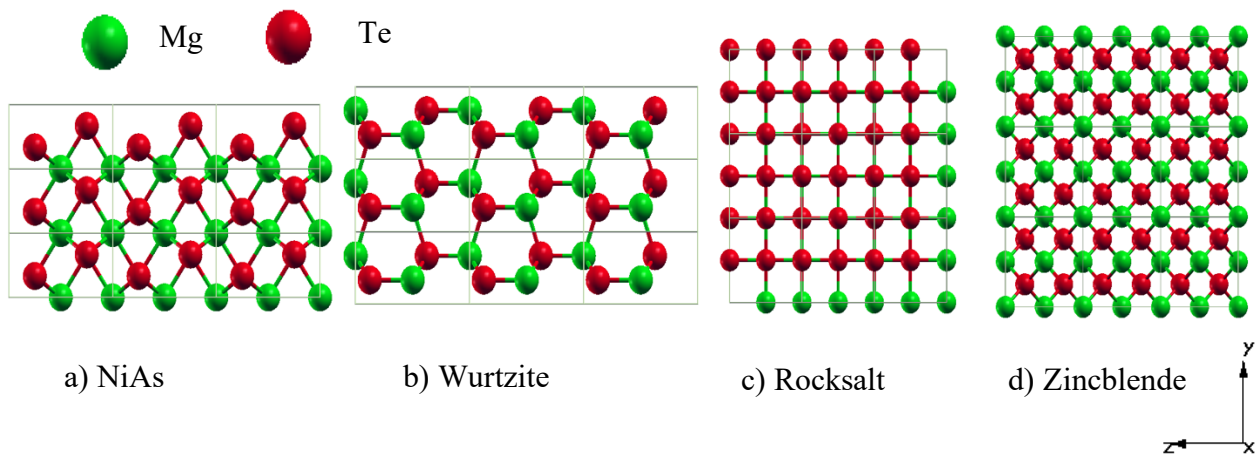

a) NiAs      b) Wurtzite      c) Rocksalt      d) Zincblende

Figure 1a-d): Crystal structure of MgTe with crystalline phases; NiAs (a=7.8585 bohr, c/a=1.6281), wurtzite (a=8.5287 bohr, c/a=1.6286), rocksalt (a=11.0985 bohr) and zincblende (a=12.073 bohr) respectively.

Lattice thermal conductivity($k$) was computed by solving phonon Boltzmann transport equation (PBTE)[37] in both single mode relaxation approximation (SMRT) and iteratively using a variational method. Expression for thermal conductivity ($k$) obtained by solving PBTE in the single mode relaxation time (SMRT) approximation[38] is given by,

$$k_\alpha = \frac{\hbar^2}{N\Omega k_b T^2} \sum_\lambda v_{\alpha\lambda}^2 \omega_\lambda^2 \bar{n}_\lambda (\bar{n}_\lambda + 1)\tau_\lambda \quad (1)$$

where, $\alpha, \hbar, N, \Omega, k_b, T$, are the cartesian direction, Planck constant, size of the **q** mesh, unit cell volume, Boltzmann constant, and absolute temperature respectively. $\lambda$ represents the vibrational mode (**q**$j$) (**q** is the wave vector and $j$ represent phonon polarization). $\omega_\lambda, \bar{n}_\lambda$, and $v_{\alpha\lambda} (= \partial\omega_\lambda/\partial q)$ are the phonon frequency, equilibrium Bose-Einstein population and group velocity along cartesian direction $\alpha$, respectively of a phonon mode $\lambda$. $\omega_\lambda, \bar{n}_\lambda$, and $c_{\alpha\lambda}$ are derived from the knowledge of phonon dispersion computed using 2$^{nd}$ order IFCs. $\tau_\lambda$ is the phonon lifetime and is computed using the equation,

$$\frac{1}{\tau_\lambda} = \pi \sum_{\lambda'\lambda''} |V_3(-\lambda, \lambda', \lambda'')|^2 \times [2(n_{\lambda'} - n_{\lambda''})\delta(\omega(\lambda) + \omega(\lambda') - \omega(\lambda'')) + (1 + n_{\lambda'} + n_{\lambda''})\delta(\omega(\lambda) - \omega(\lambda') - \omega(\lambda''))] \quad (2)$$

where, $\frac{1}{\tau_\lambda}$ is the anharmonic scattering rate based on the lowest order three phonon interactions and $V_3(-\lambda, \lambda', \lambda'')$ are the three-phonon coupling matrix elements computed using both harmonic (2$^{nd}$ derivative of energy) and anharmonic (3$^{rd}$ derivative of energy) interatomic force constants. 2$^{nd}$ and 3$^{rd}$ order interatomic force constants were derived from density-functional perturbation theory (DFPT)[39, 40]. Harmonic force constants for ZB and RS systems were calculated on 8 x 8 x 8(9 x 9 x 6 for WZ and NiAs) q-grid. Anharmonic force constants for ZB and RS were computed on a 4 x 4 x 4 (3 x 3 x 2 for WZ and NiAs) q point grid using D3Q[37, 41, 42] package within QUANTUM-ESPRESSO. Acoustic sum rules were imposed on both harmonic and anharmonic interatomic force constants. Phonon linewidth and lattice thermal conductivity were calculated using 'thermal2' package within QUANTUM ESPRESSO. For these calculations, 30 x 30 x 30(for ZB and RS) and 30 x 30 x 20(for WZ and NiAs) q -mesh was used and iterations in the exact solution of the PBTE were performed until $\Delta k$ between consecutive iterations diminished to below 1.0e$^{-5}$. $k$ values were converged after 5 iterations. Casimir scattering[43] is imposed to include the effect of boundary scattering for computing length dependent thermal conductivity in the nanoscales.

Elastic constants were computed using 'thermo_pw' package in QUANTUM-ESPRESSO; Voigt-Reuss-Hill approximation[44] was used to calculate Bulk modulus, Shear modulus(G), Young's Modulus(E) and Poisson's ratio(υ).

**Results and Discussion:** Phonon dispersion and phonon density of states for the four crystalline phases of MgTe is shown in Fig 1 which are in good agreement with previous work[3]. Structural parameters such as Young's modulus(E), Bulk modulus(B), Shear modulus(G) and Poisson's ratio computed based on Voigt-Ruess-Hill approximation are listed in Table 1 which are also in excellent agreement with the previously published work[3, 17] for all the four crystalline phases of MgTe.

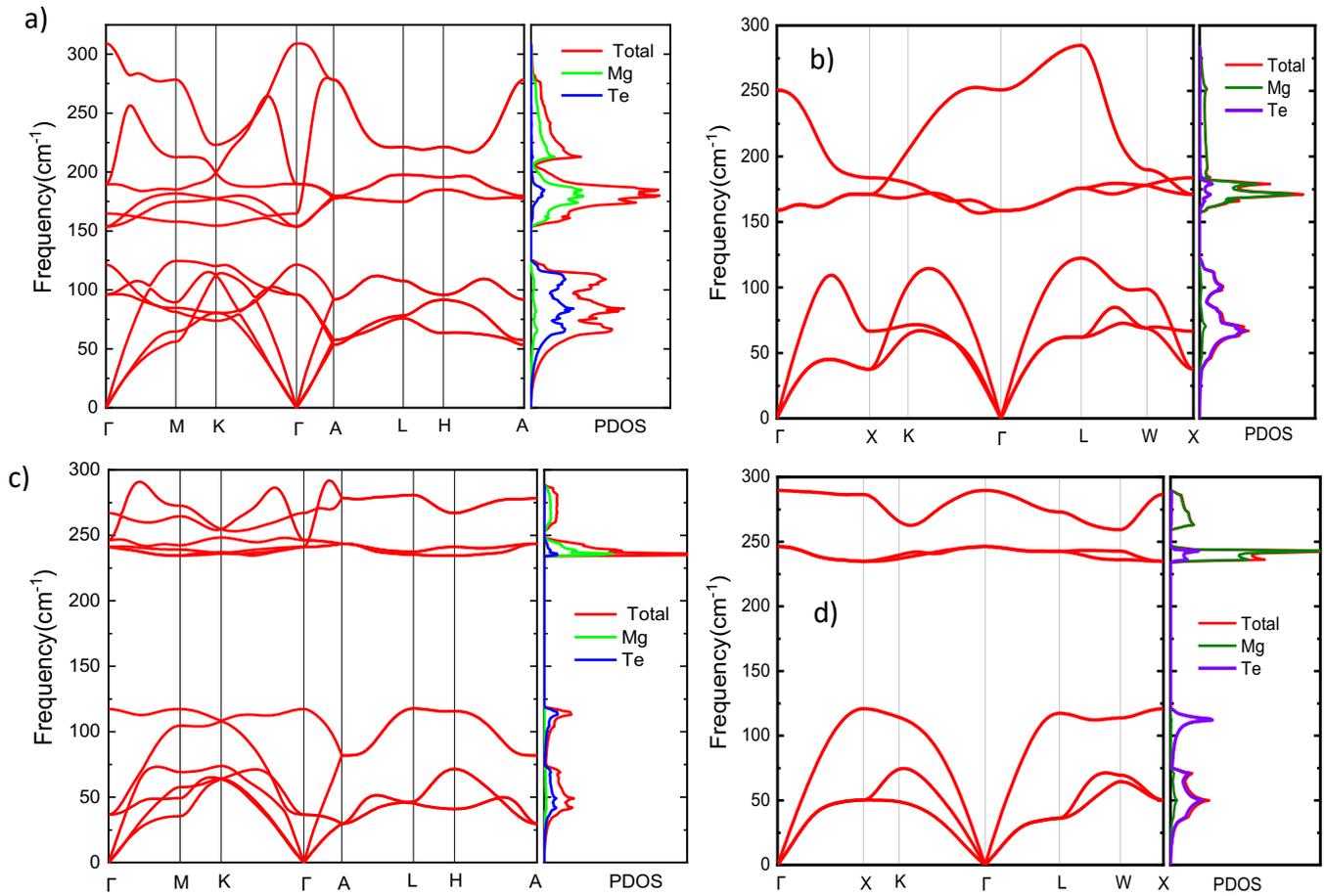

Figure 2: Phonon dispersion and phonon density of states for the MgTe with crystal phase a) nickel arsenic b) rocksalt c) wurtzite and d) zincblende

**Table 1: Lattice constants, Bulk modulus(B), Youngs modulus(E), Shear modulus(G) and poisson's(υ) ratio of MgTe with different crystal phase.**

| S. No | Crystal phase | a (bohr) | c/a | B (GPa) | E(GPa) | G(GPa) | υ |
|---|---|---|---|---|---|---|---|
| 1. | Nickel arsenic (NiAs) | 7.8585 | 1.6281 | 52.82 | 63.97 | 24.64 | 0.2981 |
| 2. | Wurtzite (WZ) | 8.5287 | 1.6286 | 38.97 | 44.92 | 17.18 | 0.3076 |
| 3. | Rocksalt (RS) | 11.0985 | | 52.7 | 86.12 | 35.07 | 0.2276 |
| 4. | Zincblende (ZB) | 12.073 | | 38.39 | 37.1 | 13.88 | 0.3367 |

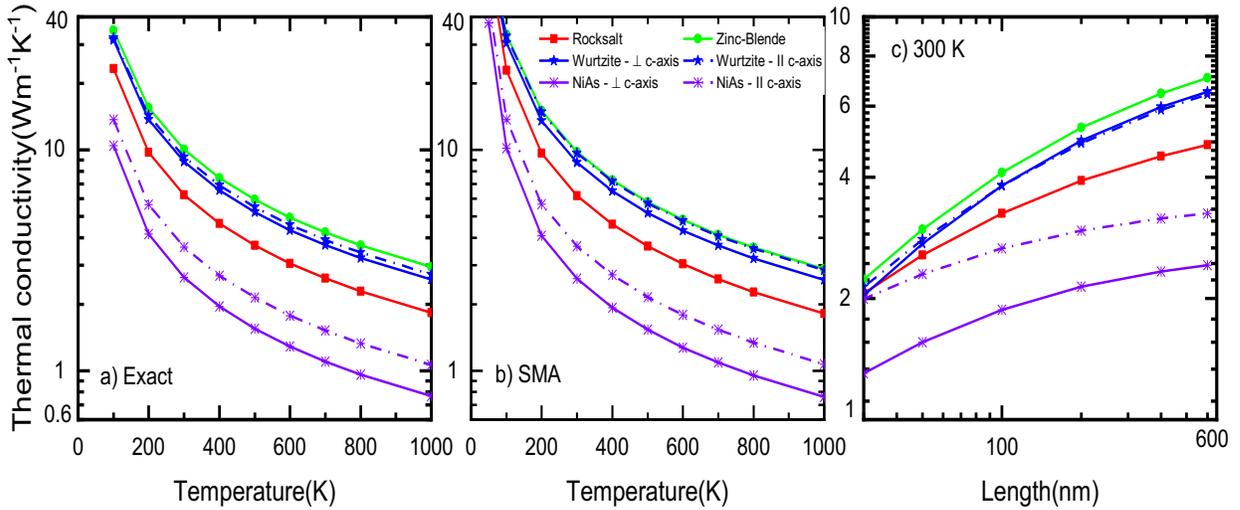

Figure 3a Temperature dependent lattice thermal conductivity by solving the PBTE iteratively b) at single mode relaxation time approximation (SMA) c) Length dependence thermal conductivity at room temperature (300 K) for MgTe with different crystalline phase.

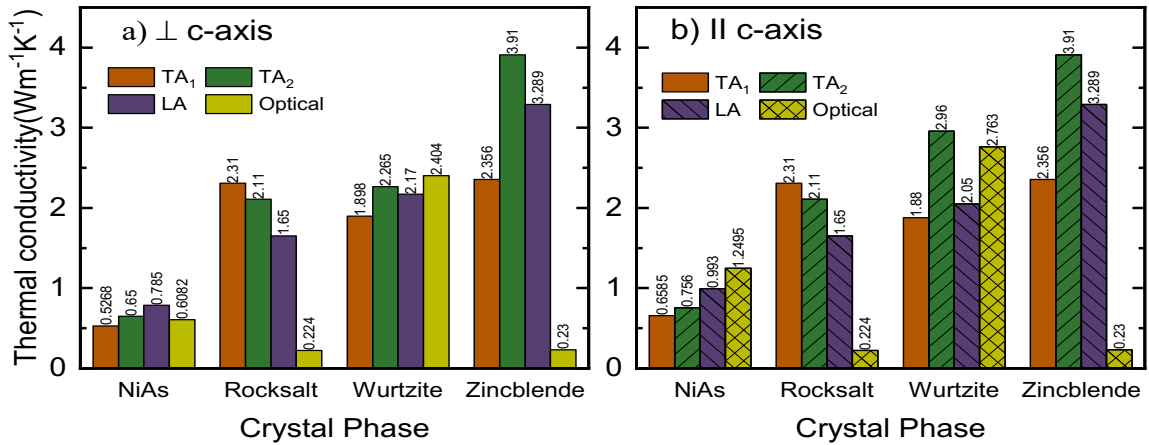

Figure 4: Thermal conductivity contribution from $TA_1$, $TA_2$, LA and optical phonon modes of MgTe with different crystalline phase along different directions.

Lattice thermal conductivity(*k*) calculated by solving the phonon Boltzmann transport equation (PBTE) is shown in Fig 3. Fig 3a and b represents the temperature dependent lattice thermal conductivity of MgTe with different crystalline phase by solving the PBTE iteratively and at SMA. SMA results are just 5% less than that of the iterative solutions. At 300 K, full iterated thermal conductivity($k$) of MgTe is as follows: $k_{NiAs}$(2.645 Wm$^{-1}$K$^{-1}$) < $k_{RS}$(6.26 Wm$^{-1}$K$^{-1}$) < $k_{WZ}$(8.83 Wm$^{-1}$K$^{-1}$) < $k_{ZB}$(10.05 Wm$^{-1}$K$^{-1}$). These low thermal conductivity of less than ~ 10 Wm$^{-1}$K$^{-1}$ shows the promising nature of MgTe in thermoelectric applications. $k$ of 10.05 Wm$^{-1}$K$^{-1}$ for the zincblende phase MgTe is 3.8 times of the $k$ of NiAs phase. This is due to the large phonon bandgap (~ 100 cm$^{-1}$) which eliminates the phonon scattering rates. Whereas ultra-low thermal conductivity of 2.645 Wm$^{-1}$K$^{-1}$ for the NiAs phase is due to the acoustic phonons are scattered by low frequency optical phonons because of the less phonon bandgap (~25 cm$^{-1}$). To explain this, we have analyzed the mode contributions thermal conductivity of transverse acoustic (TA$_1$, TA$_2$), longitudinal acoustic (LA) and optical phonon modes, phonon group velocities, phonon linewidths (scattering rates) and its spectral distribution.

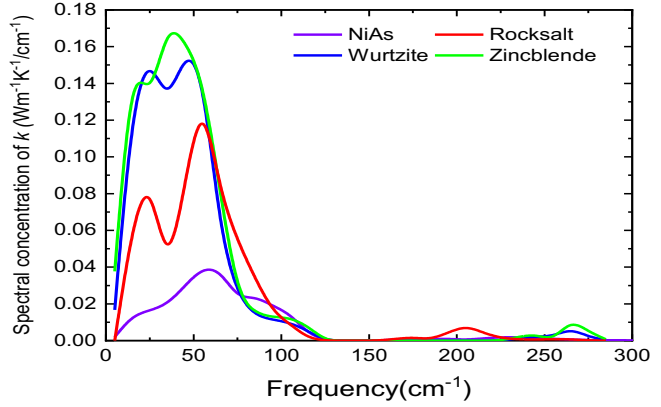

Figure 5: Spectral distributions of thermal conductivity in MgTe with different crystalline phase.

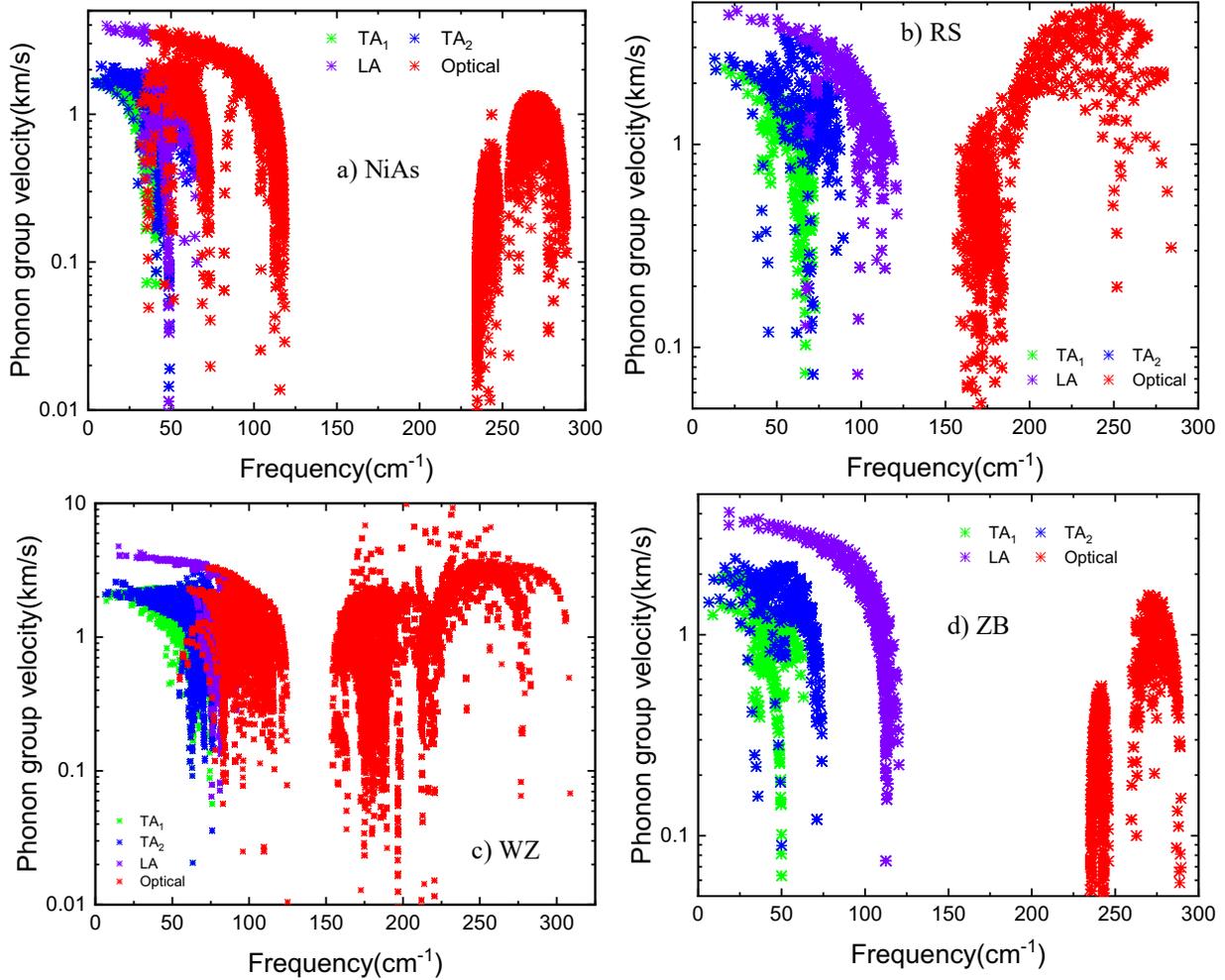

Figure 7 : Phonon group velocity of MgTe with crystalline phase a) nickel arsenic (NiAs) b) rocksalt (RS) b) wurtzite (WZ) d) zincblende (ZB) at 300K

Figure 4 represents the thermal conductivity contribution from each phonon mode along ⊥ and ∥ to the c-axis (For cubic MgTe $k$ along ⊥ - c-axis and II – c-axis are same) at single-mode relaxation time approximation. For the cubic systems, optical phonon contributions are less than ~3.5 %. Whereas optical phonons has a major contributions in both wurtzite and Nias crystal phase due to the low frequency optical phonons. For an example, 1.245 Wm$^{-1}$K$^{-1}$ along the c-axis for with NiAs phase is 34.2% to its overall thermal conductivity and is higher than both TA and LA phonon modes. Likewise, 2.404 Wm$^{-1}$K$^{-1}$ along ⊥ -c-axis is 27.5 % to its overall thermal conductivity in wurtzite MgTe. To understand this, we plotted a spectral distribution of thermal conductivity over the entire frequency (Fig. 5) and we can observe that, low frequency optical phonons has significant contributions to its overall thermal thermal conductivity. Whereas in cubic (NiAs and ZB) MgTe, TA modes between 25 - 75 cm$^{-1}$ has a major contribution to its overall thermal conductivity. To illustrate this further, we have plotted the phonon group velocities and phonon linewidths for all the crystalline phases in Fig. 6 and Fig. 7.

Fig 6 a-d represents the phonon group velocities of MgTe with different crystalline phase. We can observe that, low frequency optical phonons (less than 130 cm$^{-1}$) has a considerable phonon group velocities to that of the acoustic phonons in NiAs and wurtzite phase. Fig 7 a-d shows the phonon linewidth for MgTe with different crystalline phase. In cubic systems, zincblende has the lowest phonon linewidth (less than 2 cm$^{-1}$) for acoustic modes due to large phonon bandgap and has the highest thermal conductivity (~10 Wm$^{-1}$K$^{-1}$) whereas TA phonons in rocksalt has one order of magnitude higher scattering rate than of the zincblende and has low thermal conductivity. Likewise, TA and LA phonons in NiAs has 8 times scattering rate than its counterpart wurtzite structure. Optical phonons in NiAs has considerable phonon lifetime (inverse of scattering rate) and hence has a significant contribution to its overall thermal conductivity.

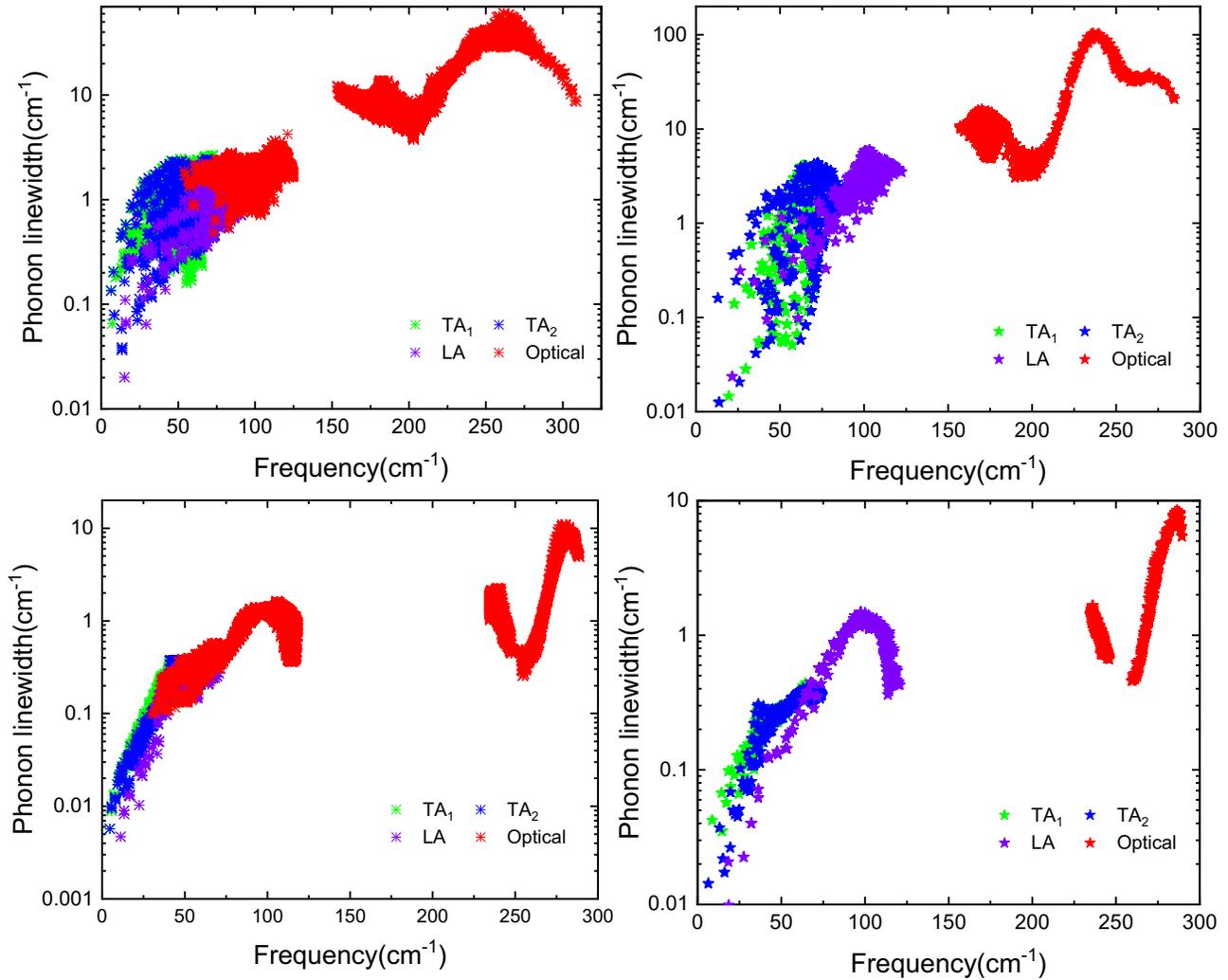

Figure 4 : Phonon linewidths of MgTe for the crystalline phase a) nickel arsenic (NiAs) b) rocksalt (RS) c) wurtzite (WZ) and zincblende (ZB) at 300K.

For the nanostructures, length dependent thermal conductivity of MgTe between 30 nm and 1000 nm is computed by introducing the boundary/Casimir scattering and is shown in Fig 3c. At 300K and at 100 nm, zincblende has a maximum thermal conductivity of ~ 4 $Wm^{-1}K^{-1}$ shows the promising nature of MgTe for the thermoelectric applications.

**Conclusion:** In this work, thermal conductivity of magnesium telluride (MgTe) with four crystalline phases; zincblende, rocksalt, wurtzite and nickel arsenic were computed by first principles calculations with phonon Boltzmann transport equations. Our first principles calculations shows a low thermal conductivity of less than ~ 10 Wm$^{-1}$K$^{-1}$ for all the crystalline phase of MgTe. We systematically investigated the phonon group velocity, phonon scattering rate and mode dependent thermal conductivity of MgTe. Our first principles calculations shows that, NiAs and wurtzite has significant contributions from optical phonons than ZB and rocksalt. At nanometer length scales such as 50 nm for NiAs phase, thermal conductivity of less than 1.4 Wm$^{-1}$K$^{-1}$ shows a promising nature of MgTe for thermoelectric applications.

**Acknowledgements:** R.M and J.G would like to acknowledge OU Supercomputing Center for Education Research (OSCER) for providing computational resources. J.G and R.M acknowledge financial support from NSF CAREER grant, Award # 1847129.